\documentstyle{article}

\font\bdi=cmmib10
\font\tenof=msym10
\def\bi#1{\hbox{\bdi #1\/}}
\def\xb{\bi{x}}
\def\zb{\bi{z}}
\def\vb{\bi{v}}
\def\zetab{\bi{\char'20}}
\def\sigmab{\bi{\char'33}}
\def\R{\mbox{\tenof R}}
\def\N{\mbox{\tenof N}}
\def\case#1#2{{\textstyle{#1\over #2}}}
\newcommand{\sumij}{\sum_{\scriptstyle i,j \atop \scriptstyle i\ne j}}
\newcommand{\sumjk}{\sum_{\scriptstyle j,k \atop \scriptstyle i\ne j\ne k\ne i}}
\newcommand{\D}{\hat D}
\newcommand{\diag}{\mathop{\rm diag}\nolimits}
\newcommand{\tP}{\tilde P}

\title{\hfill{\normalsize ULB/229/CQ/97/4}\\
\vspace{1cm}
Three-body Generalizations of the Sutherland Problem}
\author{C. Quesne\thanks{Directeur de recherches FNRS; E-mail:
cquesne@ulb.ac.be}
\\ {\small \sl Physique Nucl\'eaire Th\'eorique et Physique Math\'ematique,}\\
{\small \sl Universit\'e Libre de Bruxelles, Campus de la Plaine CP229,} \\
{\small \sl  Boulevard~du Triomphe, B-1050 Brussels, Belgium}}
\date{ }
\begin{document}
\maketitle
\begin{abstract}
The three-particle Hamiltonian obtained by replacing the two-body trigonometric
potential of the Sutherland problem by a three-body one of a similar form is
shown to be exactly solvable. When written in appropriate variables, its
eigenfunctions can be expressed in terms of Jack symmetric polynomials. The
exact
solvability of the problem is explained by a hidden $sl(3,{\bf R})$ symmetry. A
generalized Sutherland three-particle problem including both two- and three-body
trigonometric potentials and internal degrees of freedom is then
considered. It is
analyzed in terms of three first-order noncommuting differential-difference
operators, which are constructed by combining SUSYQM supercharges with the
elements of the dihedral group~$D_6$. Three alternative commuting operators are
also introduced.
\end{abstract}
%
%
\section{Introduction}
In 1974, Calogero and Marchioro~\cite{marchioro}, on one hand, and
Wolfes~\cite{wolfes}, on the other hand, extended the Calogero
problem~\cite{calogero} for three particles on a line interacting via
inverse-square
two-body potentials (and harmonic forces in the case of bound states) to a
problem
where there is an additional three-body potential of a similar form. Later
on, it
was pointed out by Olshanetsky and Perelomov~\cite{perelomov} that the
Calogero-Marchioro-Wolfes (CMW) problem is related to the root system of the
exceptional Lie algebra~$G_2$, and to the Weyl group of the latter, namely the
dihedral group~$D_6$. More recently, the Brink {\it et al\/}~\cite{brink} and
Polychronakos~\cite{poly} exchange operator formalism was extended to the
CMW~problem to deal with particles with internal degrees of freedom, thereby
leading to a $D_6$-extended Heisenberg algebra~\cite{cq95}.\par
%
%
In this communication, we present some results for similar generalizations
of the
Sutherland problem~\cite{sutherland}, wherein trigonometric potentials are
considered instead of inverse-square ones~\cite{cq96,cq97}. The Hamiltonian
considered here is
\begin{eqnarray}
  H & = & - \sum_{i=1}^3 \partial_i^2 + g a^2 \sum_{\scriptstyle i,j=1 \atop
          \scriptstyle i\ne j}^3 \csc^2 \left(a(x_i-x_j)\right) \nonumber
\\
  & & \mbox{} + 3f a^2  \sum_{\scriptstyle i,j,k=1 \atop \scriptstyle i\ne
j\ne k\ne
          i}^3 \csc^2  \left(a(x_i+x_j-2x_k)\right), \label{eq:H}
\end{eqnarray}
where $x_i$, $i=1$, 2,~3, $0\le x_i\le \pi/a$, denote the particle coordinates,
$\partial_i \equiv \partial/\partial x_i$, and $g$, $f$ are assumed not to
vanish
simultaneously and to be such that $g > -1/4$, $f > -1/4$. In the case
where $g\ne
0$ and $f = 0$, Hamiltonian~(\ref{eq:H}) reduces to the Sutherland
Hamiltonian~\cite{sutherland}, while for~$a \to 0$, it goes over into the
CMW~Hamiltonian~\cite{marchioro,wolfes}.\par
%
%
Hamiltonian~(\ref{eq:H}) is invariant under translations of the centre-of-mass,
whose coordinate will be denoted by $R = (x_1 + x_2 + x_3)/3$. In other
words, $H$
commutes with the total momentum $P = -i \sum_{i=1}^3 \partial_i$, which may be
simultaneously diagonalized. It proves convenient to use two different systems
of relative coordinates, namely $x_{ij} \equiv x_i - x_j$, $i\ne j$, and
$y_{ij} \equiv
x_i + x_j - 2x_k$, $i\ne j\ne k\ne i$, where in the latter, we suppressed
index~$k$
as it is entirely determined by $i$ and~$j$.\par
%
%
Since the potentials are singular and crossing is therefore not allowed, in
the case
of distinguishable particles the wave functions in different sectors of
configuration space are disconnected, while for indistinguishable
particles, they
are related by a symmetry requirement.\par
%
%
{}For distinguishable particles in a given sector of configuration space, the
unnormalized ground-state wave function of Hamiltonian~(\ref{eq:H}) is given by
\begin{equation}
  \psi_0(\xb) = \prod_{\scriptstyle i,j=1 \atop \scriptstyle i\ne j}^3
  \left|\sin(a x_{ij})\right|^{\kappa} \left|\sin(a y_{ij})\right|^{\lambda},
  \label{eq:gswf}
\end{equation}
where $\kappa \equiv \left(1 + \sqrt{1+4g}\right)/2$ or~0, and $\lambda \equiv
\left(1 + \sqrt{1+4f}\right)/2$ or~0, according to whether $g\ne 0$ or
$g=0$, and
$f\ne0$ or $f=0$, respectively (or, equivalently, $g = \kappa (\kappa -
1)$, $f =
\lambda (\lambda - 1)$). The corresponding eigenvalues of~$H$ and~$P$ are $E_0 =
8 a^2 (\kappa^2 + 3\kappa\lambda + 3\lambda^2)$, and $p_0=0$~\cite{cq96}.\par
%
%
In Sec.~2, we will prove that Hamiltonian~(\ref{eq:H}) with pure three-body
interactions, i.e., for~$g=0$, is exactly solvable, and we will derive its
energy
spectrum and eigenfunctions. In Sec.~3, we will propose an extension
of~(\ref{eq:H}) for a system of three particles with internal degrees of
freedom,
and introduce the corresponding exchange operator formalism.\par
%
%
\section{Exact Solvability of the Pure Three-Body Problem}
Let us assume that $g=0$ (hence $\kappa=0$), and $f\ne0$ in
Eq.~(\ref{eq:H}). For
distinguishable particles in a given sector of configuration space, the
simultaneous
solutions of the eigenvalue equations $H \psi(\xb) = E \psi(\xb)$, and $P
\psi(\xb) =
p \psi(\xb)$ can be found by setting $\psi(\xb) = \psi_0(\xb) \varphi(\xb)$. The
functions~$\varphi(\xb)$ satisfy the equations $h \varphi(\xb) =
\epsilon \varphi(\xb)$, and $P \varphi(\xb) = p \varphi(\xb)$, where $h \equiv
(\psi_0(\xb))^{-1} (H - E_0) \psi_0(\xb)$, and $\epsilon \equiv E - E_0$.
In terms of
the new variables $z_i \equiv \exp\left(\frac{2}{3} i a (x_i - 2x_j +
4x_k)\right)$, where $(ijk) = (123)$, the gauge-transformed Hamiltonian~$h$
becomes
\begin{equation}
  h = 12 a^2 \left( \sum_i \bigl(z_i \partial_{z_i}\bigr)^2 + \lambda
  \sumij \frac{z_i + z_j}{z_i - z_j} z_i \partial_{z_i}\right) - \frac{8}{3} a^2
  \biggl(\sum_i z_i \partial_{z_i}\biggr)^2,
\end{equation}
while $P = 2 a \sum_i z_i \partial_{z_i}$.\par
%
%
It can be easily proved~\cite{cq97} that the eigenfunctions and eigenvalues
of~$h$
and~$P$ are given by
\begin{equation}
  \varphi_{\{k\}}(\xb) = \exp(6i a q R) J_{\{\mu\}}\left(\zb;
\lambda^{-1}\right),
  \label{eq:eigenf}
\end{equation}
and
\begin{eqnarray}
  \epsilon_{\{ k\}} & = & 4a^2 \left[3 \sum_i k_i^2 - \case{2}{3} \left(\sum_i
           k_i\right)^2 - 6 \lambda^2\right], \nonumber \\
  p_{\{k\}} & = & 2a \sum_i k_i = 2a \left(\sum_i \mu_i + 3q\right),
           \label{eq:eigenv}
\end{eqnarray}
where $J_{\{\mu\}}\left(\zb; \lambda^{-1}\right)$ denotes the Jack  (symmetric)
polynomial in the variables $z_i$, $i=1$, 2,~3, corresponding to the
parameter~$\lambda^{-1}$, and the partition $\{\mu\} = \{\mu_1 \mu_2\}$ into not
more than two parts~\cite{stanley}. In Eqs.~(\ref{eq:eigenf})
and~(\ref{eq:eigenv}),
$k_1 = q - \lambda$, $k_2 = \mu_2 + q$, $k_3 = \mu_1 + q + \lambda$, and $q\in
\R$. In Table~\ref{tab:table1}, the explicit form of
$J_{\{\mu\}}\left(\zb;\lambda^{-1}\right)$ is given for $\mu_1+\mu_2 \le
4$.\par
%
%
\begin{table}[h]

\caption{Jack polynomials $J_{\{\mu\}}\left(\zb;\lambda^{-1}\right)$ for $\mu_1
+\mu_2 \le 4$.}
\label{tab:table1}

\vspace{0.5cm}
\begin{tabular}{ll}
  \hline\\[-0.2cm]
  $\{\mu\}$ & $J_{\{\mu\}}\left(\zb;\lambda^{-1}\right)$ \\[0.3cm]
  \hline\\[-0.2cm]
  $\{0\}$ & 1 \\[0.3cm]
  $\{1\}$ & $\sum_i z_i$ \\[0.3cm]
  $\{1^2\}$ & $\sum_{i<j} z_i z_j$ \\[0.3cm]
  $\{2\}$ & $\sum_i z_i^2 + \frac{2\lambda}{\lambda+1} \sum_{i<j} z_i z_j$
       \\[0.3cm]
  $\{21\}$ & $\sum_{i\ne j} z_i^2 z_j + \frac{6\lambda}{2\lambda+1} z_1 z_2
       z_3$ \\[0.3cm]
  $\{3\}$ & $\sum_i z_i^3 + \frac{3\lambda}{\lambda+2} \sum_{i\ne j} z_i^2 z_j
       + \frac{6\lambda^2}{(\lambda+1)(\lambda+2)} z_1 z_2 z_3$ \\[0.3cm]
  $\{2^2\}$ & $\sum_{i<j} z_i^2 z_j^2 + \frac{2\lambda}{\lambda+1} z_1 z_2 z_3
       \sum_i z_i$ \\[0.3cm]
  $\{31\}$ & $\sum_{i\ne j} z_i^3 z_j + \frac{2\lambda}{\lambda+1} \sum_{i<j}
       z_i^2 z_j^2 + \frac{\lambda(5\lambda+3)}{(\lambda+1)^2} z_1 z_2 z_3
\sum_i
       z_i$ \\[0.3cm]
  $\{4\}$ & $\sum_i z_i^4 + \frac{4\lambda}{\lambda+3}
       \sum_{i\ne j} z_i^3 z_j + \frac{6\lambda(\lambda+1)}
       {(\lambda+2)(\lambda+3)} \sum_{i<j} z_i^2 z_j^2$ \\[0.3cm]
  & $\mbox{} + \frac{12\lambda^2}{(\lambda+2)(\lambda+3)} z_1 z_2 z_3 \sum_i
       z_i$ \\[0.3cm]
  \hline
\end{tabular}
\end{table}
%
%
The eigenfunctions of~$h$ can be separated into centre-of-mass and relative
functions as follows:
\begin{equation}
  \varphi_{\{k\}}(\xb) = \exp\biggl[2i a \biggl(\sum_i k_i\biggr) R\biggr]
  P_{\{\mu\}}\left(\zetab;\lambda^{-1}\right),
\end{equation}
where $P_{\{\mu\}}\left(\zetab;\lambda^{-1}\right)$ is the polynomial in
$\zeta_1 \equiv \sum_i v_i$ and~$\zeta_2 \equiv \sum_{i<j} v_i v_j$, $v_i \equiv
\exp(-2i a x_{jk}) = z_i \exp(-2iaR)$ for $(ijk) = (123)$, obtained from the
corresponding Jack polynomial $J_{\{\mu\}}\left(\vb;\lambda^{-1}\right)$ by
making the change of variables $v_i \to \zeta_1$, $\zeta_2$. It satisfies the
eigenvalue equation
\begin{equation}
  h^{rel} P_{\{\mu\}}\left(\zetab;\lambda^{-1}\right) = \epsilon_{\{\mu\}}^{rel}
  P_{\{\mu\}}\left(\zetab;\lambda^{-1}\right),
\end{equation}
where
\begin{eqnarray}
  h^{rel} & = & 8 a^2 \biggl[ \left(\zeta_1^2 - 3\zeta_2\right)
\partial^2_{\zeta_1}
         + (\zeta_1 \zeta_2 - 9) \partial^2_{\zeta_1\zeta_2} +
\left(\zeta_2^2 - 3
         \zeta_1\right) \partial^2_{\zeta_2} \nonumber \\
  & & \mbox{} + (3\lambda + 1) \left(\zeta_1 \partial_{\zeta_1} + \zeta_2
         \partial_{\zeta_2}\right)\biggr], \label{eq:hrel}  \\
  \epsilon_{\{\mu\}}^{rel} & = & 8 a^2 \left(\mu_1^2 - \mu_1 \mu_2 + \mu_2^2
  + 3 \lambda \mu_1\right).   \label{eq:rel-energy}
\end{eqnarray}
The relative energies are similar to those obtained with pure two-body
interactions, i.e. for $g\ne 0$ and $f=0$. In Table~\ref{tab:table2}, they
are listed
for $\mu_1+\mu_2 \le 4$, together with the corresponding eigenfunctions
$P_{\{\mu\}}\left(\zetab;\lambda^{-1}\right)$. On the results displayed in the
Table, it can be checked that $P_{\{\mu\}}\left(\zetab;\lambda^{-1}\right)$
belongs
to the space $V_{\mu_1}(\zetab)$, where $V_n(\zetab)$, $n \in \N$, is defined as
the space of polynomials in $\zeta_1$ and~$\zeta_2$ that are of degree less than
or equal to~$n$ (hence, $\dim V_n = (n+1)(n+2)/2$).\par
%
%
\begin{table}[h!]

\caption{Eigenvalues~$\epsilon_{\{\mu\}}^{rel}$ and
eigenfunctions~$P_{\{\mu\}}\left(\zetab;\lambda^{-1}\right)$ of~$h^{rel}$ for
$\mu_1 + \mu_2 \le 4$.}
\label{tab:table2}

\vspace{0.5cm}
\begin{tabular}{llll}
  \hline\\[-0.2cm]
  $\{\mu\}$ & $\epsilon_{\{\mu\}}^{rel}/(8a^2)$ &
  $P_{\{\mu\}}\left(\zetab;\lambda^{-1}\right)$
  \rule[-0.3cm]{0cm}{0.6cm}\\[0.3cm]
  \hline\\[-0.2cm]
  $\{0\}$ & 0 & 1 \\[0.3cm]
  $\{1\}$ & $3\lambda+1$ & $\zeta_1$ \\[0.3cm]
  $\{1^2\}$ & $3\lambda+1$ & $\zeta_2$ \\[0.3cm]
  $\{2\}$ & $2(3\lambda+2)$ & $\zeta_1^2 - \frac{2}{\lambda+1} \zeta_2$
\\[0.3cm]
  $\{21\}$ & $3(2\lambda+1)$ & $\zeta_1 \zeta_2 - \frac{3}{2\lambda+1}$
       \\[0.3cm]
  $\{3\}$ & $9(\lambda+1)$ & $\zeta_1^3 - \frac{6}{\lambda+2} \zeta_1 \zeta_2
       + \frac{6}{(\lambda+1)(\lambda+2)}$ \\[0.3cm]
  $\{2^2\}$ & $2(3\lambda+2)$ & $\zeta_2^2 - \frac{2}{\lambda+1} \zeta_1$
       \\[0.3cm]
  $\{31\}$ & $9\lambda+7$ & $\zeta_1^2 \zeta_2 - \frac{2}{\lambda+1} \zeta_2^2
       - \frac{3\lambda+1}{(\lambda+1)^2} \zeta_1$ \\[0.3cm]
  $\{4\}$ & $4(3\lambda+4)$ & $\zeta_1^4 - \frac{12}{\lambda+3} \zeta_1^2
       \zeta_2 + \frac{12}{(\lambda+2)(\lambda+3)} (\zeta_2^2 + 2\zeta_1)$
       \\[0.3cm]
  \hline
\end{tabular}
\end{table}
%
%
In Ref.~\cite{cq97}, the degeneracies of the relative energy
spectrum~(\ref{eq:rel-energy}) were obtained for both distinguishable and
indistinguishable (either bosonic or fermionic) particles on the line
interval $(0,
\pi/a)$, interacting via pure two-body or three-body potential. It was shown
that although the results do not depend upon the nature of interactions for
distinguishable particles, they do for indistinguishable ones. Such a
property is due
to the fact that both the configuration space sectors, and the variables
the relative
wave functions depend upon have different transformation properties under
particle permutations for the problems with pure two-body or pure three-body
potential.\par
%
%
The exact solvability of~$H$ for $g=0$ and $f\ne0$, or equivalently
of~$h^{rel}$,
defined in Eq.~(\ref{eq:hrel}), can be easily explained by a hidden $sl(3,\R)$
symmetry~\cite{cq97}. The Hamiltonian~$h^{rel}$ can indeed be rewritten as a
quadratic combination
\begin{eqnarray}
  h^{rel} & = & 8 a^2 \bigl[E_{11}^2 + E_{11} E_{22} + E_{22}^2 - 3 E_{12}
E_{32}
         - 3 E_{21} E_{31} \nonumber \\
  & & \mbox{} - 9 E_{31} E_{32} + 3 \lambda \left(E_{11} + E_{22}\right)\bigr]
\end{eqnarray}
of the operators
\begin{eqnarray}
  E_{11} & = & \zeta_1 \partial_{\zeta_1}, \qquad E_{22} = \zeta_2
            \partial_{\zeta_2}, \qquad E_{33} = n - \zeta_1 \partial_{\zeta_1}
            - \zeta_2 \partial_{\zeta_2}, \nonumber \\
  E_{21} & = & \zeta_2 \partial_{\zeta_1}, \qquad E_{12} = \zeta_1
            \partial_{\zeta_2}, \nonumber \\
  E_{31} & = & \partial_{\zeta_1}, \qquad E_{13} = n \zeta_1 - \zeta_1^2
            \partial_{\zeta_1} - \zeta_1 \zeta_2 \partial_{\zeta_2},
\nonumber \\
  E_{32} & = & \partial_{\zeta_2}, \qquad E_{23} = n \zeta_2  - \zeta_1 \zeta_2
            \partial_{\zeta_1} - \zeta_2^2 \partial_{\zeta_2},
\end{eqnarray}
satisfying $gl(3,\R)$ commutation relations $\left[E_{ij}, E_{kl}\right] =
\delta_{kj} E_{il} - \delta_{il} E_{kj}$, together with the constant trace
condition
$\sum_i E_{ii} = n$ for any real $n$~value. Whenever $n$~is a non-negative
integer,
the operators~$E_{ij}$ preserve the space $V_n(\zetab)$. Hence, $h^{rel}$
preserves
an infinite flag of spaces, $V_0(\zetab) \subset V_1(\zetab) \subset V_2(\zetab)
\subset \ldots$. Its representation matrix is therefore triangular in the basis
wherein all spaces $V_n(\zetab)$ are naturally defined, so that $h^{rel}$
is exactly
solvable. This result is similar to that previously obtained for the pure
two-body
trigonometric potential~\cite{turbiner}.\par
%
%
\section{Two- and Three-body Problem with Internal Degrees of Freedom}
Let us now assume that both $g$ and~$f$ are nonvanishing. From the ground-state
wave function~(\ref{eq:gswf}) of Hamiltonian~(\ref{eq:H}), one can construct
SUSYQM supercharge operators $\hat Q^+$, $\hat Q^- = \bigl(\hat
Q^+\bigr)^{\dagger}$, whose matrix elements can be expressed in terms of six
differential operators $Q_i^{\pm} = \mp \partial_i - \partial_i \ln
\psi_0(\xb)$,
$i=1$, 2,~3~\cite{andrianov}. The latter are given by~\cite{cq96}
\begin{eqnarray}
  Q^{\pm}_i & = & \mp \partial_i - \kappa a \sum_{j\ne i} \cot(a x_{ij})
\nonumber
           \\
  & & \mbox{} - \lambda a \left(\sum_{j\ne i} \cot(a y_{ij}) - \sumjk \cot(a
           y_{jk})\right).
\end{eqnarray}
The corresponding supersymmetric Hamiltonian is $\hat H = \diag(H^{(0)},
H^{(1)},$  $H^{(2)}, H^{(3)})$, where $H^{(0)} = H - E_0 =\sum_i Q^+_i
Q^-_i$, $H^{(1)}$
and $H^{(2)}$ contain matrix potentials, while $H^{(3)} = \sum_i Q^-_i
Q^+_i$ only
differs from~$H^{(0)}$ by the replacement in~$H$ of $g = \kappa (\kappa -
1)$, $f =
\lambda (\lambda - 1)$ by $g = \kappa (\kappa + 1)$, $f = \lambda (\lambda
+ 1)$,
respectively.\par
%
%
In the case of the CMW~problem, it was shown in Ref.~\cite{cq95} that the
corresponding operators~$Q^-_i$ can be transformed into three commuting
differential-difference operators~$D_i$, the so-called Dunkl operators of the
mathematical literature~\cite{dunkl}, by inserting in appropriate places some
finite-group elements $K_{ij}$ and $L_{ij} \equiv K_{ij} I_r$. Here
$K_{ij}$ are particle permutation operators, while $I_r$ is the inversion
operator in
relative-coordinate space. In the centre-of-mass coordinate system, they satisfy
the relations
\begin{eqnarray}
  K_{ij} & = & K_{ji} = K_{ij}^{\dagger}, \quad K_{ij}^2 = 1, \quad K_{ij}
K_{jk} =
          K_{jk} K_{ki} = K_{ki} K_{ij}, \nonumber \\
  K_{ij} I_r & = & I_r K_{ij}, \quad I_r = I_r^{\dagger}, \quad I_r^2 = 1,
\nonumber \\
  K_{ij} x_j & = & x_i K_{ij}, \quad K_{ij} x_k = x_k K_{ij}, \quad I_r x_i
= - x_i
          I_r,   \label{eq:D6}
\end{eqnarray}
for all $i\ne j\ne k\ne i$. The operators 1, $K_{ij}$, $K_{ijk} \equiv
K_{ij} K_{jk}$,
$I_r$, $L_{ij}$, and $L_{ijk} \equiv K_{ijk} I_r$, where $i$, $j$, $k$ run
over the set
\{1, 2, 3\}, are the 12 elements of the dihedral group~$D_6$.\par
%
%
By proceeding in a similar way in the present problem, we find the three
differential-difference operators~\cite{cq96}
\begin{eqnarray}
  D_i & = & \partial_i - \kappa a \sum_{j\ne i} \cot(a x_{ij}) K_{ij}
\nonumber \\
  & & \mbox{} - \lambda a \left(\sum_{j\ne i} \cot(a y_{ij}) L_{ij} -
\sumjk \cot(a
         y_{jk}) L_{jk}\right),
\end{eqnarray}
where $i=1$, 2,~3. From their definition and Eq.~(\ref{eq:D6}), it is
obvious that
such operators are both anti-Hermitian and $D_6$-covariant, i.e.,
$D_i^{\dagger} = - D_i$, $K_{ij} D_j = D_i K_{ij}$, $K_{ij}D_k = D_k
K_{ij}$, and $I_r
D_i = - D_i I_r$, for all $i\ne j\ne k\ne i$, but that they do not commute among
themselves. Their commutators are indeed given by
\begin{equation}
  \left[D_i, D_j\right] = - a^2 \left(\kappa^2 + 3 \lambda^2 - 4 \kappa
\lambda I_r
  \right) \sum_{k\ne i,j} \left(K_{ijk} - K_{ikj}\right), \qquad i\ne j,
\end{equation}
and only vanish in the $a\to 0$ limit, i.e., for the CMW~problem.\par
%
%
The operators~$D_i$ may be used to construct a generalized Hamiltonian with
exchange terms
\begin{eqnarray}
  H_{exch} & \equiv & - \sum_i \partial_i^2 + a^2 \sumij \csc^2(a x_{ij}) \kappa
         (\kappa - K_{ij}) \nonumber \\
  & & \mbox{} + 3 a^2 \sumij \csc^2(a y_{ij}) \lambda(\lambda - L_{ij})
         \nonumber \\
  & = & -\sum_i D_i^2 + 6 a^2 \left(\kappa^2 + 3 \lambda^2\right) \nonumber \\
  & & \mbox{} + a^2 \left(\kappa^2 + 3 \lambda^2 + 12 \kappa \lambda I_r\right)
         \left(K_{123} + K_{132}\right).   \label{eq:Hexch}
\end{eqnarray}
In those subspaces of Hilbert space wherein $\left(K_{ij}, L_{ij}\right) =
(1,1)$,
$(1,-1)$, $(-1,1)$, or~$(-1,-1)$, the latter reduces to Hamiltonian~(\ref{eq:H})
corresponding to $(g,f) = (\kappa (\kappa-1), \lambda (\lambda-1))$, $(\kappa
(\kappa-1), \lambda (\lambda+1))$, $(\kappa (\kappa+1), \lambda
(\lambda-1))$, or
$(\kappa (\kappa+1), \lambda (\lambda+1))$, respectively.\par
%
%
{}From the operators~$D_i$ and the elements $K_{ij}$, $L_{ij}$ of~$D_6$, it
is also
possible to construct an alternative set of three Dunkl operators, i.e., three
anti-Hermitian, commuting, albeit non-covariant, differential-difference
operators
\begin{eqnarray}
  \D_i & = & D_i + i \kappa a \sum_{j\ne i} \alpha_{ij} K_{ij} + i \lambda
a \left(
         \sum_{j\ne i} \beta_{ij} L_{ij} - \sumjk \beta_{jk} L_{jk}\right),
\nonumber
         \\
  & & \alpha_{ij}, \beta_{ij} \in \R,
\end{eqnarray}
in terms of which the generalized Hamiltonian with exchange terms, defined in
Eq.~(\ref{eq:Hexch}), can be rewritten as $H_{exch} = - \sum_i \D_i^2$. By
choosing
for the $\alpha_{ij}$'s the values previously considered for the Dunkl
operators of
the pure two-body problem~\cite{bernard}, namely $\alpha_{ij} = - \alpha_{ji} =
-1$, $i<j$, one finds~\cite{cq96} that there are four equally acceptable
choices for
the remaining constants $\beta_{ij}$: $(\beta_{12}, \beta_{23}, \beta_{31}) =
(-1,1,1)$, $(-1,1,-1)$, $(- 5/3, 1/3, 1/3)$, and $(- 1/3, 5/3, - 1/3)$.\par
%
%
The transformation properties under~$D_6$ of the new operators~$\D_i$ are given
by
\begin{eqnarray}
  K_{ij} \D_j - \D_i K_{ij} & = & - i \kappa a \Biggl(2 \alpha_{ij} +
\sum_{k\ne i,j}
       \left(\alpha_{ik} - \alpha_{jk}\right) K_{ijk}\Biggr) \nonumber \\
  & & - i \lambda a \sum_{k\ne i,j} \left(\beta_{ik} - \beta_{jk}\right) I_r
       \left(K_{ijk} + 2 K_{ikj}\right), \nonumber \\
  & &  i\ne j, \nonumber \\
  \left[K_{ij}, \D_k\right] & = & i a \left[\kappa \left(\alpha_{ik} -
\alpha_{jk}
       \right) - \lambda \left(\beta_{ik} - \beta_{jk}\right) I_r\right]
\left(K_{ijk}
       - K_{ikj}\right), \nonumber \\
  & & i\ne j\ne k\ne i, \nonumber \\
  \left\{I_r, \D_i\right\} & = & 2i \kappa a \sum_{j\ne i} \alpha_{ij} L_{ij}
       \nonumber \\
       & & \mbox{} + 2i \lambda a \left(\sum_{j\ne i} \beta_{ij} K_{ij} - \sumjk
       \beta_{jk} K_{jk}\right).
\end{eqnarray}
\par
%
%
The Hamiltonian with exchange terms~$H_{exch}$ can be related to a Hamiltonian
${\cal H}^{(\kappa,\lambda)}$ describing a one-dimensional system of three
particles with $SU(n)$ ``spins'' (or colours in particle physics language),
interacting via spin-dependent two and three-body potentials~\cite{cq96},
\begin{eqnarray}
  {\cal H}^{(\kappa,\lambda)} & = & - \sum_i \partial_i^2 + a^2 \sumij \csc^2(a
           x_{ij}) \kappa (\kappa - P_{ij}) \nonumber \\
  & & \mbox{} + 3 a^2 \sumij \csc^2(a y_{ij}) \lambda (\lambda - \tilde
P_{ij}).
\end{eqnarray}
Here each particle is assumed to carry a spin with $n$ possible values, and
$P_{ij}$, $\tilde P_{ij} \equiv P_{ij} \tilde P$ are some operators acting
only in spin
space. The operator~$P_{ij}$ is defined as the operator permuting the $i$th and
$j$th spins, while $\tP$ is a permutation-invariant and involutive
operator, i.e.,
$\tP \sigma_i = \sigma_i^* \tP$, for some $\sigma^*_i$ such that
$P_{jk} \sigma_i^* = \sigma_i^* P_{jk}$ for all $i$, $j$,~$k$, and
$\sigma_i^{**} =
\sigma_i$. For $SU(2)$ spins for instance, $\sigma_i = \pm 1/2$, $P_{ij} =
(\sigma^a_i \sigma^a_j + 1)/2$, where
$\sigma^a$, $a=1$, 2,~3, denote the Pauli matrices, $\tP$ may be taken as 1 or
$\sigma^1_1 \sigma^1_2 \sigma^1_3$, and accordingly $\sigma_i^* =
\sigma_i$ or~$-\sigma_i$. The operators $P_{ij}$ and~$\tP$ satisfy
relations similar to those fulfilled by $K_{ij}$ and~$I_r$ (cf.\
Eq.~(\ref{eq:D6})),
with $x_i$ and $-x_i$ replaced by $\sigma_i$ and $\sigma_i^*$ respectively.
Hence
1, $P_{ij}$, $P_{ijk} \equiv P_{ij} P_{jk}$, $\tP$, $\tilde P_{ij}$, and
$\tilde P_{ijk}
\equiv P_{ijk} \tilde P$ realize the dihedral group~$D_6$ in spin space. Such a
realization will be referred to as $D_6^{(s)}$ to distinguish it from the
realization
$D_6^{(c)}$ in coordinate space, corresponding to $K_{ij}$ and~$I_r$.\par
%
%
The Hamiltonian ${\cal H}^{(\kappa,\lambda)}$ remains invariant under the
combined
action of~$D_6$ in coordinate and spin spaces (to be referred to as
$D_6^{(cs)}$),
since it commutes with both $K_{ij} P_{ij}$ and $I_r \tP$. Its eigenfunctions
corresponding to a definite eigenvalue therefore belong to a (reducible or
irreducible) representation of~$D_6^{(cs)}$. For indistinguishable
particles that are
bosons (resp.~fermions), only those irreducible representations of
$D_6^{(cs)}$ that
contain the symmetric (resp.~antisymmetric) irreducible representation of the
symmetric group $S_3$ should be considered. There are only two such inequivalent
representations, which are both one-dimensional and denoted by $A_1$ and $B_1$
(resp.~$A_2$ and $B_2$)~\cite{hamermesh}. They differ in the eigenvalue of $I_r
\tP$, which is equal to $+1$ or $-1$, respectively.\par
%
%
In such representations, for an appropriate choice of the parameters
$\kappa$,~$\lambda$, ${\cal H}^{(\kappa,\lambda)}$ can be obtained from
$H_{exch}$ by applying some projection operators. Let indeed $\Pi_{B\pm}$
(resp.~$\Pi_{F\pm}$) be the projection operators that consist in replacing
$K_{ij}$
and $I_r$ by $P_{ij}$ (resp.~$-P_{ij}$) and $\pm \tP$, respectively, when they
are at the right-hand side of an expression. It is obvious that
$\Pi_{B\pm}(H_{exch}) = {\cal H}^{(\kappa,\pm\lambda)}$, and
$\Pi_{F\pm}(H_{exch}) = {\cal H}^{(-\kappa,\pm\lambda)}$. If $H_{exch}$ has been
diagonalized on a basis of functions depending upon coordinates and spins,
then its
eigenfunctions $\Psi(\xb,\sigmab)$ are also eigenfunctions of ${\cal
H}^{(\kappa,\pm\lambda)}$ (resp.~${\cal H}^{(-\kappa,\pm\lambda)}$)
provided that
$(K_{ij} - P_{ij}) \Psi(\xb,\sigmab) = 0$ (resp.\ $(K_{ij} + P_{ij})
\Psi(\xb,\sigmab)
= 0$) and $(I_r \mp \tP) \Psi(\xb,\sigmab) = 0$. \par
%
%
In conclusion, the three-body generalization of the Sutherland problem with
internal
degrees of freedom, corresponding to the Hamiltonian~${\cal
H}^{(\kappa,\lambda)}$, is directly connected with the corresponding problem
with exchange terms, governed by the Hamiltonian~$H_{exch}$. The exchange
operator formalism developed for the latter should therefore be relevant to a
detailed study of the former.\par
%
%

\end{document}